\documentclass[aps,prb,reprint,twocolumn,superscriptaddress,showpacs,numerical]{revtex4-1}
\usepackage{amsmath,bm}
\usepackage{graphicx,color}
\usepackage{times}
\def\vs{\vec s}
\def\bra#1{\langle{#1}|}
\def\ket#1{|{#1}\rangle}
\def\ave#1{\langle{#1}\rangle}
\def\hc{\rm H.c.}
\def\const{\rm const.}
\DeclareMathOperator{\tr}{tr}
\begin{document}
\title{Manipulation of two spin qubits in a double quantum dot using an electric field}
\date{\today}
\author{Atsuo Shitade}
\affiliation{Department of Applied Physics, The University of Tokyo, Hongo, Bunkyo-ku, Tokyo 113-8656, Japan}
\author{Motohiko Ezawa}
\affiliation{Department of Applied Physics, The University of Tokyo, Hongo, Bunkyo-ku, Tokyo 113-8656, Japan}
\author{Naoto Nagaosa}
\affiliation{Department of Applied Physics, The University of Tokyo, Hongo, Bunkyo-ku, Tokyo 113-8656, Japan}
\affiliation{Cross-Correlated Materials Research Group (CMRG), and Correlated Electron Research Group (CERG), RIKEN-ASI, Wako 351-0198, Japan}
\begin{abstract}
We propose purely electric manipulation of spin qubits 
by means of the spin-orbit interaction (SOI) {\it without 
magnetic field or magnets} in a double quantum dot.
All the unitary transformations can be constructed by the 
time-dependent Dzyaloshinsky-Moriya interaction 
between the two spins, which arises from the Rashba SOI modulated by electric field.
As a few demonstrations, we study both analytically and numerically the three operations, i.e., 
(A) the spin initialization,
(B) the two-spin rotation in the opposite directions,
and (C) the two-spin rotation in the same direction.
The effects of the relaxation and the feasibility of this proposal are also discussed. 
\end{abstract}
\maketitle
\section{Introduction}
Manipulation of spins in semiconductors is a subject of extensive studies both 
theoretically and experimentally. Especially, the possible application of the 
electron spins to the quantum computations attracts much attention,
and the control of a single spin or two spins in quantum dot systems
aiming at the qubit operations with large-scale integration is an important issue.
Loss and DiVincenzo~\cite{PhysRevA.57.120} proposed the implementation of the universal set of quantum gates 
by using the time-dependent exchange interaction and local magnetic field.
It is known that the {\tt XOR} and the single-spin operations are enough to construct 
any quantum computations.~\cite{PhysRevA.52.3457}
The {\tt SWAP} operation $U_{\rm SWAP}$ and its square root $U_{\rm SWAP}^{1/2}$ 
can be realized by the exchange interaction with a certain period of time,
and the quantum {\tt XOR} gate by the combination of $U_{\rm SWAP}^{1/2}$ and the single-spin rotations 
induced by magnetic field or a ferromagnet.
Another proposal is the electron spin resonance transistors in Si-Ge with the $g$-factor 
modulated by the electric field serving the possible qubit system.~\cite{PhysRevA.62.012306}

Experimentally, the gate voltage can control electron spins of a double quantum dot,
which contains initialization, manipulation, and read-out.~\cite{J.R.Petta09302005}
With the help of magnetic field, the singlet and one of the triplet form the two-level system,
in which the {\tt SWAP} operation and the singlet-triplet spin echo have been
demonstrated.~\cite{J.R.Petta09302005}

However, it is desirable to control spins purely electrically since it is difficult 
to apply magnetic field confined in a nanoscale region.
It was proposed to use the decoherence-free subspace, 
in which the exchange interaction alone is universal.~\cite{PhysRevLett.85.1758}
Soon later, it was proposed that a single qubit can be encoded by 
three spins, where only two in eight ($= 2^3$) quantum states are used.~\cite{Nature.408.339}
In this proposal, the global magnetic field is inevitable in initialization.

\begin{figure}
  \centering
  \includegraphics[clip,width=0.35\textwidth]{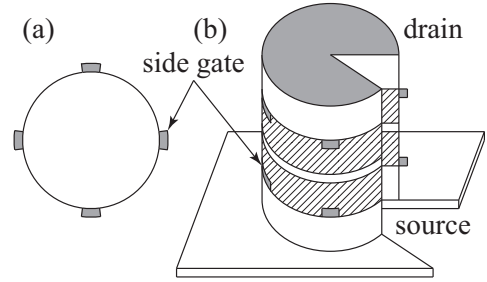}
  \caption{%
  The schematic (a) top and (b) sectional views of the vertical double quantum dot we propose.
  Shaded and gray regions indicate quantum dots and electrodes.
  Four side gates are attached to each dot, which makes it possible to apply electric field in three directions.
  }
  \label{fig:setup}
\end{figure}

In this paper, we propose fully electric manipulation of spins in a double quantum dot,
in which magnetic field is not necessary at all even in initialization.
Furthermore, in our method, two-bit operation is realized only by 
using a double quantum dot in contrast to the previous proposal.~\cite{Nature.408.339} 
We explicitly show the universal set of quantum gates can be constructed by the exchange and
the Dzyaloshinsky-Moriya (DM) interactions.~\cite{Dzyaloshinsky1958241,PhysRev.120.91}

Most of the electric manipulation methods of spins employ the relativistic spin-orbit 
interaction (SOI),~\cite{PhysRevB.74.165319,PhysRevLett.97.240501,PhysRevB.77.235301,PhysRevA.81.022315}
and it has been already demonstrated that the Rashba SOI 
can be controlled by the gate voltage in GaAs system.~\cite{PhysRevLett.78.1335}
The Rashba interaction is written as 
\[
  H_{\rm R}
  = \lambda {\vec p} \cdot \vs \times {\vec E}
\]
with ${\vec p}$, $\vs$ being the momentum and spin of an electron, respectively,  while ${\vec E}$ is electric field and $\lambda$ is the Rashba coupling constant.
In a double quantum dot, the Rashba SOI in the region between the two dots leads to the spin rotation associated with the transfer of the electron, i.e., 
\[
  H_{\rm T}
  = -t c_1^{\dagger} e^{i{\vec \theta} \cdot \vs/2} c_2 + \hc,
\]
where $c_i^{\dagger} = (c_{i\uparrow}^{\dagger}, c_{i\downarrow}^{\dagger})$ is the creation operator of the electron at $i$th dot, $t$ is the transfer integral, 
and $e^{i{\vec \theta} \cdot \vs/2}$ is the ${\rm SU}(2)$ matrix corresponding to the spin 
rotation around the axis $ {\vec \theta} \parallel {\vec e}_{12} \times {\vec E}$ with 
${\vec e}_{12}$ the unit vector connecting the sites 1 and 2.  
When spin $1/2$ is localized at each dot,
the transfer integral together with the Coulomb charging energy leads to 
the DM interaction [with ${\vec D} \parallel {\vec \theta}$ in Eq.~\eqref{eq:ham}]
as discussed in semiconductor nanostructures.~\cite{PhysRevB.64.075305,PhysRevB.69.075302}
Note that the DM interaction is the two-spin interaction, 
and it appears difficult to manipulate only a single spin with use of the DM interaction
since the single-spin Hamiltonian necessarily breaks the time-reversal symmetry.
In the ingenious method using the DM interaction and the spin anisotropy, the two-spin encoding scheme was adopted,
i.e., two of four states in every nearest-neighboring pair of spins are used.~\cite{PhysRevLett.93.140501}
Also it is shown that the SOI together with only one component of magnetic field can manipulate a single-spin qubit.~\cite{PhysRevB.74.165319,PhysRevLett.97.240501}
In contrast, we will show below that
it is possible to construct the single-spin operations using the time-dependent DM interaction without any magnetic field or magnetic anisotropy.

\section{Perturbation Theory}
\begin{figure}
  \centering
  \includegraphics[clip,width=0.25\textwidth]{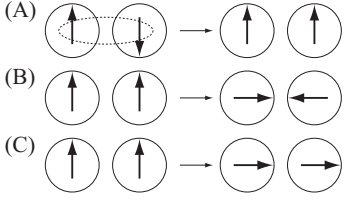}
  \caption{
  Schematics of the two-qubit operations.
  (A) Preparation of fully polarized triplet state as the initialization from the singlet 
      ground state,
  (B) the two-spin rotation in the opposite directions,
  and (C) the two-spin rotation in the same direction.
  Black circles indicate quantum dots, and the black broken oval indicates the 
  singlet entanglement due to the exchange interaction $J$.%
  }
  \label{fig:operation}
\end{figure}

Below we explicitly construct the unitary transformations from the 
exchange and DM interactions,
\begin{equation}
  H(t)
  = H_0 + H^{\prime}(t)
  = J(\vs_1 \cdot \vs_2 - 1/4) + {\vec D}(t) \cdot \vs_1 \times \vs_2.
  \label{eq:ham}
\end{equation}
Here the exchange interaction $J$ is assumed to be constant,
and the DM vector ${\vec D}(t)$ to have three components,
which breaks the overall axial symmetry.
For the realization of this Hamiltonian, we propose the vertical double quantum dot 
schematically shown in Fig.~\ref{fig:setup}.
With this configuration, four side gates attached to each dot make it possible
to produce the Rashba SOI and to shift the centers of wave functions in the dots separately.
The former leads to the DM interaction,
and the latter is essential to make the $z$ component of the DM interaction.~\cite{tarucha}
These two issues are both determined by the spatial symmetry,
i.e. the original $C_4$ and some mirror symmetries with respect the $xy$ plane is lowered by the electric field.
We show that temporal changes of ${\vec D}(t)$ enable
(A) to initialize spins from the singlet ground state 
to one of the triplet states,
(B) to rotate two spins in the opposite directions,
and (C) to rotate two spins in the same direction,
which are schematically shown in Fig.~\ref{fig:operation}.
Combining (B) and (C), we can construct the single-spin rotation. 

First of all, we investigate the time-evolution operator $U(t)$ within the perturbation
theory in the DM interaction since it is usually smaller than the exchange coupling $J$.
Up to the first-order perturbation in $H^{\prime}$, it is given by
\begin{equation}
  U^{(1)}(t)
  = e^{-iH_0t}\exp\left[-i\int_0^t{\rm d}t_1H_{\rm I}^{\prime}(t_1)\right],
  \label{eq:uni1}
\end{equation}
in which $H_{\rm I}^{\prime}(t) = e^{iH_0t}H^{\prime}(t)e^{-iH_0t}$ denotes the perturbation Hamiltonian in the interaction picture,
and we put $\hbar = 1$.
Note that the time-ordered product $T$ in front of the exponential operator is absent in the first-order approximation.
To obtain $H_{\rm I}^{\prime}(t)$, we solve the equation of motion.
Two spins obey
\[
  \frac{{\rm d}\vs_{1\rm I}(t)}{{\rm d}t}
  = -\frac{{\rm d}\vs_{2\rm I}(t)}{{\rm d}t}
  = -J\vs_{1\rm I}(t) \times \vs_{2\rm I}(t),
\]
leading to
\[\begin{split}
  \frac{{\rm d}}{{\rm d}t}[\vs_{1\rm I}(t) \times \vs_{2\rm I}(t)]
  = & \frac{J}{2}(\vs_{1\rm I}(t) - \vs_{2\rm I}(t)) \\
  \frac{1}{2}\frac{{\rm d}}{{\rm d}t}[\vs_{1\rm I}(t) - \vs_{2\rm I}(t)]
  = & -J\vs_{1\rm I}(t) \times \vs_{2\rm I}(t).
\end{split}\]
Hence we can explicitly obtain
\begin{align}
  H_{\rm I}^{\prime}(t)
  = & {\vec D}(t) \cdot \left[\vs_1 \times \vs_2\cos Jt + \frac{1}{2}(\vs_1 - \vs_2)\sin Jt\right]
  \label{eq:hamint1} \\
  = & \frac{1}{2}e^{-iJt}\left[2^{-1/2}D^-(t)\ket{0}\bra{1} + iD^z(t)\ket{0}\bra{2}\right.
  \notag \\
  & \left.+ 2^{-1/2}D^+(t)\ket{0}\bra{3}\right] + \hc,
  \label{eq:hamint2}
\end{align}
where $D^{\pm}(t) = D^y(t) \pm iD^x(t)$,
and $\ket{0} = (\ket{\uparrow\downarrow} - \ket{\downarrow\uparrow})/\sqrt{2}$ is the singlet state,
while $\ket{1} = \ket{\uparrow\uparrow}$, $\ket{2} = (\ket{\uparrow\downarrow} + \ket{\downarrow\uparrow})/\sqrt{2}$,
and $\ket{3} = \ket{\downarrow\downarrow}$ are the triplet states.
It is noted here that the matrix elements of the DM interaction
connect the singlet and a linear combination of the triplet states. 
Considering that the Hilbert space of the triplet states has four real degrees of freedom
($= 2 \times 3 - 2$ corresponding to the normalization condition and the overall phase factor),
the three real coefficients ${\vec D}$  of the DM interaction appear to be not enough.
This is true if we assume that ${\vec D}$ is independent of time, but once we design the time dependence of ${\vec D}(t)$,
we can connect the singlet $\ket{0}$ with any linear combination of $\ket{1}$, $\ket{2}$, and $\ket{3}$.   

\begin{figure}
  \centering
  \includegraphics[clip,width=0.48\textwidth]{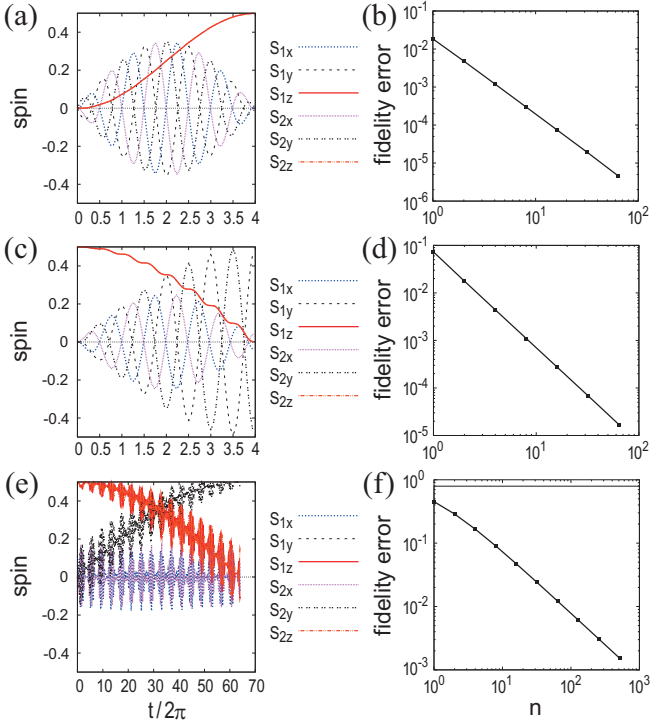}
  \caption{%
  (Color online)
  The left panels show time evolution of the spin expectation values,
  in which we choose $n = 4$ and $D/J = 0.177$ for (a), $n = 4$ and $D/J = 0.25$ for (c), and $n = 16$ and $D/J = 0.199$ for (e).
  The right ones show $n$ dependence of the fidelity errors defined in Eq.~\eqref{eq:fidelity}.
  The top two panels (a) and (b) represent initialization,
  the middle ones, (c) and (d) represent the two-spin rotation around the $x$ axis by $\pm \pi/2$, 
  leading to $\ave{\vs_{1/2}} \parallel \pm{\hat y}$,
  and the bottom ones (e) and (f) represent the two-spin rotations around the $x$ axis by $\pi/2$ corresponding to $\ave{\vs_{1/2}} \parallel {\hat y}$.
  The relaxation effect is not included.%
  }
  \label{fig:norelax}
\end{figure}

Let us consider initialization $\ket{0} \to \ket{1}$.
For this purpose we take ${\vec D}_{\rm ini}(t) = D(-\sin Jt, \cos Jt, 0)$ and put $t_0 = 2\pi/J$ in Eq.~\eqref{eq:hamint2},
which makes the first factor $e^{-iH_0t}$ in Eq.~\eqref{eq:uni1} unity, i.e., $e^{iJt_0}=1$ leading to
\[
  U_{\rm ini} = [U^{(1)}(t_0)]^n
  = \exp\left[-i\frac{n\pi D}{\sqrt{2}J}(\ket{0}\bra{1} + \ket{1}\bra{0})\right].
\]
When we start from the singlet state, this is a simple two-level problem,
\[
  U_{\rm ini}\ket{0}
  = \cos\frac{n\pi D}{\sqrt{2}J}\ket{0} - i\sin\frac{n\pi D}{\sqrt{2}J}\ket{1}.
\]
Integer $n$ is determined by the condition $n\pi D/\sqrt{2}J = \pi/2$, which gives $-i\ket{1}$.
This is asymptotically exact in the limit of $D/J \to 0$ and $n \to \infty$,
and we show in Figs.~\ref{fig:norelax}(a) and \ref{fig:norelax}(b) the numerical results for finite $n$
by taking all the higher order terms solving the equation of motion of the density matrix $\rho(t)$. 
The quantum states show oscillatory behavior in the time scale corresponding to the exchange interaction $J$,
and approach to the predicted states from the first-order perturbation as $n$ increases, in other words, as $D/J$ decreases.
It is shown that the fidelity error defined as
\begin{equation}
  1 - \left(\tr\sqrt{ \sqrt{\rho_{\infty}} \rho \sqrt{\rho_{\infty}} }\right)^2
  \label{eq:fidelity}
\end{equation}
is already of the order of $10^{-3}$ for $n = 4$.
Here $\rho$ is the calculated density matrix of the final state, i.e., $t=nt_0=\sqrt{2}\pi/D$,
while $\rho_\infty$ is that of the desired pure state shown in the right-hand side of the arrows in Fig.~\ref{fig:operation}.
As seen from Fig.~\ref{fig:norelax}(b), the fidelity error is proportional to $1/n^2$ as expected. 
This means that the conversion between the singlet and triplet can be manipulated very effectively by the DM interaction.

Next we construct the two-spin rotation in the opposite directions.
Let us consider the rotation around the $x$ axis since the same applies to that around the $y$ and $z$ axes.
Here we take ${\vec D}_{\rm opp}^x(t) = (-D\sin Jt, 0, 0)$ in Eq.~\eqref{eq:hamint1}.
Then we get
\begin{equation}
  U_{\rm opp}^{x}(\theta)
  = [U^{(1)}(t_0)]^n
  = \exp\left[i\frac{n\pi D}{2J}(s_1^x - s_2^x)\right],
  \label{eq:uopp1}
\end{equation}
which rotates two spins around the $x$ axis by $\pm \theta = \pm n\pi D/2J$.
The results of the numerical simulation for finite $n$ 
with $n\pi D/2J$ fixed at $\pi/2$ are shown in Figs.~\ref{fig:norelax}(c) and \ref{fig:norelax}(d).
Again the fidelity error is less than $10^{-2}$ even for $n = 4$, which decreases as $\propto 1/n^2$.    

The two-spin rotation in the {\it same} direction is the most nontrivial. 
We define the unitary transformations
\[
  U^{\alpha}(t_0)
  = \exp\left[i\frac{\pi D}{2J}(2(\vs_1 \times \vs_2)^{\alpha} \cos \varphi 
   - (s_1^{\alpha} - s_2^{\alpha}) \sin \varphi )\right]
\]
achieved by ${\vec D}^{\alpha}(t) = -D{\hat x}^{\alpha}\cos(Jt + \varphi)$ (${\hat x}^{\alpha} = x, y, z$)
up to the first order in $D/J$.
When we take the ``magic angle'' $\varphi = \pi/6$, we obtain 
the following composite operator from $U^{\alpha}$'s
\begin{align}
  U_{\rm same}^x(\theta)
  = & \left[U^zU^yU^{-z}U^{-y}\right]^n
  \notag \\
  \cong & \exp\left[in\left(\frac{\pi D}{2J}\right)^2(s_1^x + s_2^x)\right],
  \label{eq:usame1}
\end{align}
in which we apply the Baker-Campbell-Hausdorff formula
\[
  e^{i\lambda A}e^{i\lambda B}e^{-i\lambda A}e^{-i\lambda B}
  = \exp[-\lambda^2[A, B] + O(\lambda^3)].
\]
This is two-spin rotation around the $x$ axis by $\theta = n(\pi D/2J)^2$.
The rotation around the $y$ and $z$ axes can be obtained by cyclic permutation.
In deriving Eq.~\eqref{eq:usame1}, it is necessary that the second-order correction cancels. 
Up to the second-order perturbation, Eq.~\eqref{eq:uni1} is modified to  
\[\begin{split}
  U^{(2)}(t)
  = & e^{-iH_0t}\exp\left[-i\int_0^t{\rm d}t_1H_{\rm I}^{\prime}(t_1)\right. \\
  & \left.- \frac{1}{2}\int_0^t{\rm d}t_1
  \int_0^{t_0}{\rm d}t_2[H_{\rm I}^{\prime}(t_1), H_{\rm I}^{\prime}(t_2)] \right].
\end{split}\]
The second-order correction, which is proportional to
\[\begin{split}
  & \int_0^{2\pi}{\rm d}u_1\int_0^{u_1}{\rm d}u_2\cos (u_1 + \varphi)\cos (u_2 + \varphi)\sin (u_1 - u_2) \\
  = & \frac{\pi}{4}(1 - 2\cos 2\varphi),
\end{split}\]
exactly vanishes at $\varphi = \pi/6$.
As the result, Eq.~\eqref{eq:usame1} holds up to the second-order perturbation.
Namely, it is asymptotically exact in the limit of $D/J \to 0$ and $n \to \infty$
with $n(D/J)^2$ being fixed finite.
Figures~\ref{fig:norelax}(e) and \ref{fig:norelax}(f) show the numerical results for the finite $n$ cases.
The fidelity error in Fig.~\ref{fig:norelax}(f) decreases slowly as $\propto 1/n$,
and therefore the condition is much more stringent than that for Figs.~\ref{fig:norelax}(b) and \ref{fig:norelax}(b(d).

\begin{figure}
  \centering
  \includegraphics[clip,width=0.48\textwidth]{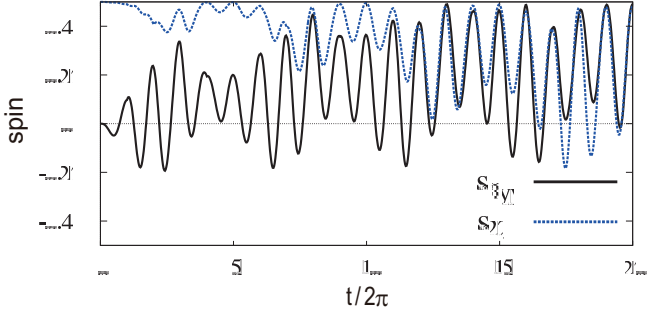}
  \caption{%
  (Color online)
  The time evolution of the spin expectation values in rotation of spin $1$ around the $x$ axis by $\pi/2$.
  We set $\theta = n\pi D/2J = \pi/4$ with $n = 4$ for the operation (B),
  and then $\theta = n(\pi D/2J)^2 = \pi/4$ with $n = 4$ for the operation (C).
  Relaxation is not included.
  }
  \label{fig:single}
\end{figure}

By combining Eqs.~\eqref{eq:uopp1} and \eqref{eq:usame1}, we can implement any single-spin rotations.
In Fig.~\ref{fig:single}, we demonstrate rotation of spin $1$ around the $x$ axis by $\pi/2$
with use of $U_{\rm opp}^x(\pi/4)$ and $U_{\rm same}^x(\pi/4)$.
The former needs the $x$ component of the DM vector, while the latter needs the $y$ and $z$ components.
Thus three components are required to implement single-spin rotations around one axis. 

The read-out process can be also designed by the DM interaction, which connects the singlet 
and triplet states. 
This is achieved by the transformation from $(N_1, N_2)=(1, 1)$ to $(0, 2)$ 
to measure the singlet probability $P_0$. 
Here $N_i=0, 1$ be the electron number at $i$th dot.
One can always measure the probability of singlet $P_0$ by 
shifting the gate voltage to transform $(N_1, N_2) = (1, 1)$ to $(0, 2)$.~\cite{J.R.Petta09302005}
By tuning the time-dependence of ${\vec D}(t)$,
one can make the situation that the singlet $\ket{0}$ is coupled to an arbitrary linear 
combination of the triplet states, called $\ket{\rm T}$.
The time evolution within this two-dimensional Hilbert space spanned by $\ket{0}$ and  $\ket{\rm T}$
can exchange the component of $\ket{0}$ and $\ket{\rm T}$ by some rotation of the angle $\pi$.
From $P_0$ after this rotation, one can measure the probability $P_{\rm T}$ of $\ket{\rm T}$.

\section{Effect of Relaxation}
We choose the integer $n$ to satisfy $n(D/J) = \const$ for (A) and (B),
and $n(D/J)^2 = \const$ for (C) to set angles.
There is a trade-off because the DM interaction $D/J$ is assumed to be small to validate the perturbation theory,
while we cannot set larger $nt_0$ than the relaxation time.
To substantiate this consideration by explicit calculation, we employ the standard boson-bath model to the relaxation.
The spin-boson interaction is described as
\[
  H_{\rm int}
  = \lambda ({\vec b}_1 \cdot \vs_1 + {\vec b}_2 \cdot \vs_2),
\]
in which
\[
  b_i^{\alpha}
  = \sum_ng_{i\alpha n}(a_{i\alpha n}^{\dagger} + a_{i\alpha n})
\]
is a fluctuating field from nuclear spins.
Here $a_{i\alpha n}$ is a bosonic operator
whose motion obeys the Hamiltonian
\[
  H_{\rm bath}
  = \sum_{i\alpha n}\omega_{i\alpha n}a_{i\alpha n}^{\dagger}a_{i\alpha n}.
\]
After the Born-Markov approximation and dropping irrelevant terms, we obtain the equation of motion of the density matrix as
\[\begin{split}
  & \frac{{\rm d}\rho_{\rm I}(t)}{{\rm d}t}
  = -{\hat \Gamma}\rho_{\rm I}(t) \\
  \equiv & -\int_0^{\infty}{\rm d}u \tr_{\rm bath}[H_{\rm intI}(t), [H_{\rm intI}(t - u), \rho_{\rm I}(t) \otimes \rho_{\rm bathI}]], \\
\end{split}\]
or
\[
  {\dot \rho}(t) = -i[H(t), \rho(t)] - {\hat \Gamma}\rho(t)
\]
in the Schr{\" o}dinger picture.
To evaluate ${\hat \Gamma}\rho$, we use
\[\begin{split}
  & \int_0^{\infty}{\rm d}u \tr_{\rm bath}b_i^{\alpha}(t)b_j^{\beta}(t - u)\rho_{\rm bath} e^{-i\omega u} \\
  = & \sum_{mn} g_{i\alpha n}g_{j\beta m} \int_0^{\infty}{\rm d}u
  \tr_{\rm bath}(a_{i\alpha n}^{\dagger}e^{i\omega_{i\alpha n}t} + a_{i\alpha n}e^{-i\omega_{i\alpha n}t}) \\
  & \times (a_{j\beta m}^{\dagger}e^{i\omega_{j\beta m}(t - u)} + a_{j\beta m}e^{-i\omega_{j\beta m}(t - u)})\rho_{\rm bath} e^{-i\omega u} \\
  = & \delta_{ij}\delta_{\alpha\beta} \sum_n {g_{i\alpha n}}^2 \int_0^{\infty}{\rm d}u \\
  & \times [n(\omega_{i\alpha n})e^{i\omega_{i\alpha n}u} + (1 + n(\omega_{i\alpha n}))e^{-i\omega_{i\alpha n}u}]e^{-i\omega u} \\
  = & \delta_{ij}\delta_{\alpha\beta} \sum_n \pi {g_{i\alpha n}}^2 \\
  & \times [n(\omega_{i\alpha n})\delta(\omega - \omega_{i\alpha n}) + (1 + n(\omega_{i\alpha n}))\delta(\omega + \omega_{i\alpha n})] \\
  = & \delta_{ij}\delta_{\alpha \beta}[n(\omega)A_{i\alpha}(\omega) + (1 + n(-\omega))A_{i\alpha}(-\omega)],
\end{split}\]
in which we define the boson distribution function $n(\omega) = (e^{\beta\omega} - 1)^{-1}$ with the inverse temperature $\beta$, and
\[
  A_{i\alpha}(\omega)
  = \sum_n \pi{g_{i\alpha n}}^2\delta(\omega - \omega_{i\alpha n})
\]
is the bath spectral function.
After straightforward calculations, we get
\[\begin{split}
  ({\hat \Gamma}\rho)_{00}
  = & \lambda^2[3n_1A_1\rho_{00} - (1 + n_1)A_1(\rho_{11} + \rho_{22} + \rho_{33})] \\
  ({\hat \Gamma}\rho)_{11/33}
  = & \lambda^2[(1 + n_1)A_1\rho_{11/33} - n_1A_1\rho_{00}] \\
  & + \lambda^2(1 + 2n_0)A_0(\rho_{11/33} - \rho_{22}) \\
  ({\hat \Gamma}\rho)_{22}
  = & \lambda^2[(1 + n_1)A_1\rho_{22} - n_1A_1\rho_{00}] \\
  & + \lambda^2(1 + 2n_0)A_0(2\rho_{22} - \rho_{11} - \rho_{33}) \\
  ({\hat \Gamma}\rho)_{01-03}
  = & \lambda^2[(1/2 + 2n_1)A_1 + (1 + 2n_0)A_0]\rho_{01-03} \\
  ({\hat \Gamma}\rho)_{12/23}
  = & \lambda^2[(1 + n_1)A_1 + 2(1 + 2n_0)A_0]\rho_{12/23} \\
  & - \lambda^2(1 + 2n_0)A_0\rho_{23/12} \\
  ({\hat \Gamma}\rho)_{13}
  = & \lambda^2[(1 + n_1)A_1 + 3(1 + 2n_0)A_0]\rho_{13}
\end{split}\]
in the singlet-triplet basis.
Here $n_0 = n(0)$ and $n_1 = n(J)$ while $A_0 = A(0)$ and $A_1 = A(J)$.
For simplicity, we neglect the directional dependence of the spectral function, i.e., $A_{i\alpha} = A$.

\begin{figure}
  \centering
  \includegraphics[clip,width=0.48\textwidth]{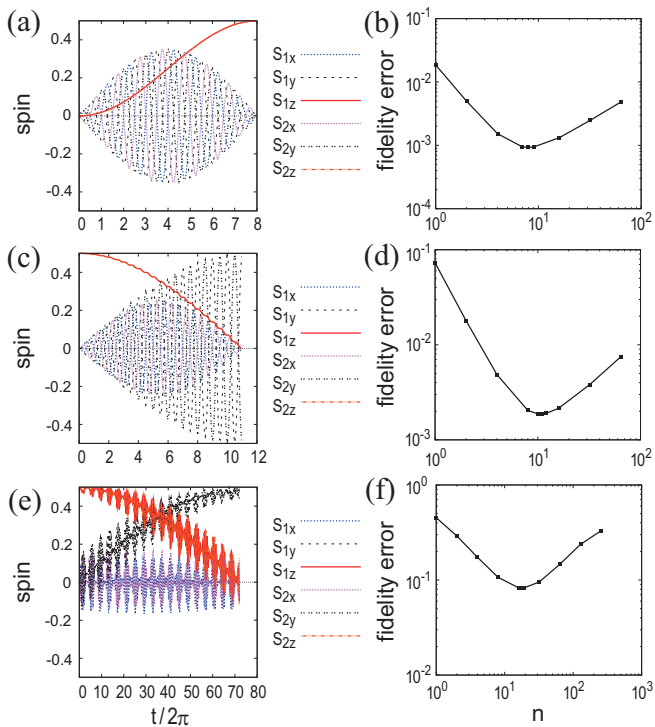}
  \caption{%
  (Color online)
  The relaxation effect is taken into account for $(2JT_1)^{-1} = (JT_2)^{-1} = 1.0 \times 10^{-4}$.
  The used parameters are $n = 8$ and $D/J = 8.84 \times 10^{-2}$ for (a),
  $n = 11$ and $D/J = 9.09 \times 10^{-2}$ for (c),
  and $n = 18$ and $D/J = 0.188$ for (e),
  which give the minimum fidelity errors, respectively. See the caption of Fig.~\ref{fig:norelax} for the notation.
  }
  \label{fig:relax}
\end{figure}

Figure~\ref{fig:relax} shows the numerical results similar to Fig.~\ref{fig:norelax} but with the relaxation. 
In Figs.~\ref{fig:relax}(b), \ref{fig:relax}(d), and \ref{fig:relax}(f), the fidelity errors have minimum at a certain $n$
because the contribution from relaxation is $\propto n$
while the discretization error decreases as $\propto 1/n^2$ or $\propto 1/n$.
Especially the two-spin rotation in the same direction is greatly affected
since it contains four unitary transformations and its fidelity error due to the higher-order perturbation decreases as $O(n^{-1})$.
To neglect the effect of relaxation, the relaxation time $T$ must be longer than $4nt_0 \sim (J/D)^2 \times (1/J)$, leading to $JT \gg (J/D)^2$.

\section{Discussion}
Finally we discuss the realistic setup of our proposal in semiconductors.
The most serious problem is the relaxation of electron spins.
The origin of relaxation is mainly the hyperfine interaction with nuclear spins.
The effective field $\hbar \gamma_{\rm e} B_{\rm nuc}$ is typically $50 {\rm neV}$ (Ref.~\onlinecite{PhysRevLett.97.056801})
and the relative magnitude $J/\hbar \gamma_{\rm e} B_{\rm nuc}$ is a crucial parameter to control the relaxation time.
In a GaAs double quantum dot, this value is $3$-$10$, and the singlet correlation decays on a time scale $10{\rm ns}$.~\cite{PhysRevLett.97.056801}
On the other hand, in the case of the singlet-triplet relaxation time of the two-electron system in a single quantum dot,
where the splitting is about $1 {\rm meV}$ and much larger than $\hbar \gamma_{\rm e} B_{\rm nuc}$, 
the hyperfine interaction is not effective, and hence the $T_1 \sim T_2 \sim 0.2 {\rm ms}$.~\cite{Nature.419.278} 
Therefore it is essential to increase $J$ and to increase $T_1$, $T_2$. 
An encouraging theoretical analysis
gives an estimate of $J \sim 1 {\rm meV}$ in coupled quantum dots,~\cite{PhysRevB.59.2070} and also the 
vertical quantum dots as shown in Fig.~\ref{fig:setup} might enhance $J$ compared with the 
horizontal dots. 

Another problem is the order of magnitude of the DM interaction generated by the electric field.
According to Nitta {\it et al}.,~\cite{PhysRevLett.78.1335}
the Rashba constant can be modified from $\lambda = 0.64 \times 10^{-11}{\rm eV m}$ at the gate voltage $V_{\rm g} = 1.5{\rm V}$
to $\lambda = 0.92 \times 10^{-11}{\rm eV m}$ at $V_{\rm g} = -1.0{\rm V}$.
This suggests that it can be modified by $1.12{\rm meVnm}$ per $1{\rm V}$.
With the typical distance between two dots used in Ref.~\onlinecite{PhysRevB.59.2070}, $2d = 28{\rm nm}$,
the SOI is estimated as $\Delta = 2(\pi/2d)\times 1.12 = 0.25{\rm meV}$ at $V_{\rm g} = 1{\rm V}$.
The ratio $D/J \sim \Delta/t$ can reach the order of $0.1$.

Under strong electric field, we may worry about the break-down phenomenon.
However, charge transfer between two dots hardly occurs except at the resonance
since the energy levels of dots are discrete.
In addition, the typical time scale of charge transfer is of the order of $100{\rm ns}$ according to Ref.~\cite{J.R.Petta09302005},
which is much longer than that of the operations we discuss, $nt_0 \sim 0.5{\rm ns}$.
Therefore we expect that charge transfer can be neglected.

In summary, we have proposed the purely electric manipulation of 
qubits in the double quantum dots in terms of the Rashba and
DM interactions which are modulated
by the time-dependent voltages. This idea might be useful also
for the control of macroscopic magnetization in the dilute 
magnetic semiconductor, which is an issue left for future
investigations.

\begin{acknowledgements}
  We are grateful to S. Tarucha, T. Otsuka, and Y. Shikano for fruitful discussions.
  A.~S. was supported by Grant-in-Aid for JSPS Fellows.
  This work was supported in part by Grant-in-Aid for Scientific Research
  (Grants No.~20940011, No.~19019004, No.~19048008, No.~19048015, No.~21244053, and No.~22740196)
  from the Ministry of Education, Culture, Sports, Science and Technology of Japan,
  Strategic International Cooperative Program (Joint Research Type) from Japan Science and Technology Agency,
  and Funding Program for World-Leading Innovative RD on Science and Technology (FIRST Program).
\end{acknowledgements}

\end{document}